\documentclass[referee]{raa}            
\usepackage{graphicx,times}            
\usepackage{lscape}
\usepackage{amsmath}
\usepackage{textcomp}
\usepackage[colorlinks,
            linkcolor=cyan,
            anchorcolor=black,
            citecolor=cyan
            ]{hyperref}
\usepackage{natbib}

\usepackage{graphicx}
\usepackage{subfigure}
\usepackage{multirow}

\begin{document}

   \title {An Investigation of Collisions between Fiber Positioning Units in LAMOST}


   \volnopage{Vol.0 (200x) No.0, 000--000}      
   \setcounter{page}{1}          

   \author{Xiao-Jie Liu
      \inst{1,2}
   \and Gang Wang
      \inst{1}
   }

   \institute{Key Laboratory of Optical Astronomy, National Astronomical Observatories, Beijing 100012, China\\; {\it liuxiaojie10@mails.ucas.ac.cn; wg@lamost.org}\\
        \and
             University of Chinese Academy of Sciences, Beijing 100049, China\\
   }

\abstract{ The arrangement of the fiber positioning units in LAMOST focal plane may lead to the collisions during the fiber allocation. To avoid these collisions, the soft protection system has to abandon some targets located in the overlapped field of the adjacent fiber units. In this paper, we firstly analyzed the probability of the collisions between fibers and inferred their possible reasons. It is useful to solve the problem of the fiber-positioning units collisions so as to improve LAMOST efficiency. Based on it, a collision handling system is designed by using the master-slave control structure between the micro control unit (MCU) and the microcomputer. The simulated experiments validate that the system can provide real-time inspection and swap the information between the fiber  unit controllers and the main controller.
}

   \authorrunning{X.-J. Liu, et al.}            
   \titlerunning{An Investigation of Fiber Collisions}  

   \maketitle

%

\section{Introduction}           
\label{sect:intro}
The Large Sky Area Multi-Object Fiber Spectroscopic Telescope (LAMOST), which is also called the Guo Shou Jing Telescope, is a special reflecting Schmidt telescope with large aperture and wide field of view. The available large focal surface accommodates up to 4000 fibers, by which the collected light of distant and faint celestial objects down to 20.5 magnitudes is fed into the spectrographs, promising a very high spectrum acquiring rate of several ten-thousands of spectra per night \citep{cui2012large,2014SPIE.9149E..1NC}.

\begin{figure}

\begin{minipage}[h]{0.5\linewidth}
\centering
\includegraphics[width=60mm,bb=0 0 500 350]{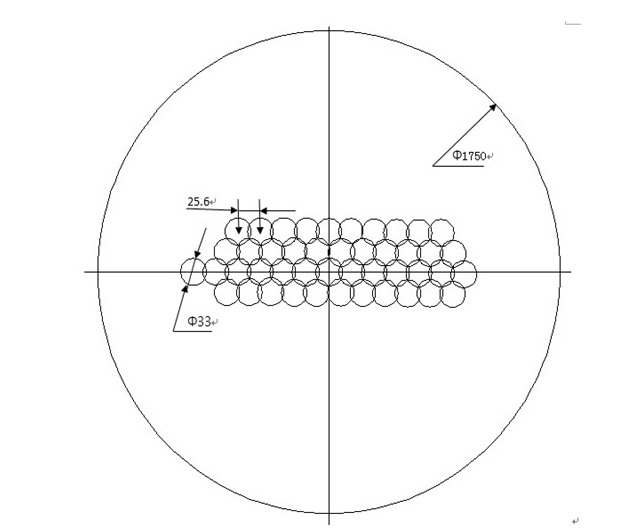}
\caption{The distribution of fiber-positioning units in the focal plate \citep{cui2012large}.}
\label{fig:side:h}
\end{minipage}%
\begin{minipage}[h]{0.5\linewidth}
\centering
\includegraphics[width=60mm,bb=0 0 500 350]{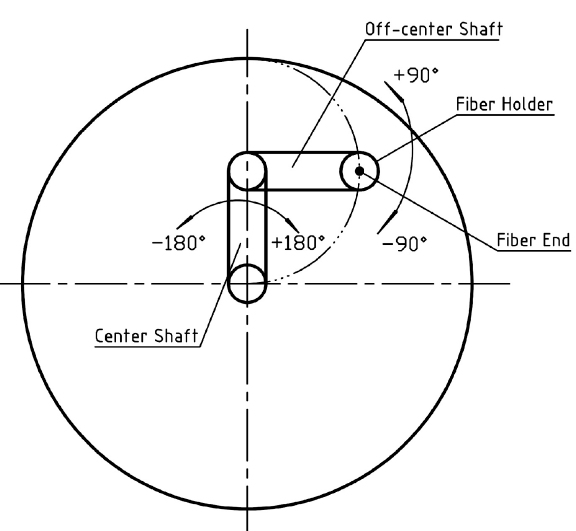}
\caption{Double revolving fiber-positioning unit \citep{cui2012large}.}
\label{fig:side:a}
\end{minipage}%
\end{figure}

The focal surface of the telescope, which has a diameter of 1.75 m, is divided into 4,000 individual domains as shown in Fig. \ref{fig:side:h}. The diameter of each fiber is 33 mm and the distance between two adjacent units is 25.6 mm, so the overlapped domain will appear in the observation field of every two adjacent fiber units \citep{cui2012large}. Therefore, the fiber arrangement can efficiently avoid the blind areas of FoVs. At the same time, the positioning procedure of the relevant fiber units may also be confused when the targets allocate in the overlapped field.
On the other hand, each unit is driven by two stepping motors \citep{xing1998parallel}, the principle of which can be seen in Fig. \ref{fig:side:a}. The unit contains a central shaft which revolves around its fixed end in a range of $\pm$180$^\circ$, and an off-center shaft which revolves around its end that is connected to the free end of the central shaft in arrange of $\pm$90$^\circ$ \citep{Liweimin+2000}. Theoretically speaking, mechanical collisions among the fiber positioning units must be avoided by the Survey Strategy System (SSS) of LAMOST which would abandon several objects.

 In this study, we firstly analyzed the probability of the collisions  among fibers by utilizing binomial probability distribution. And then, we inferred several relevant reasons for the losing targets during the observation. Based on it, a collision handling system is designed by using the master-slave control structure between the micro control unit (MCU) and the microcomputer. In order to validate its usefulness and correctness, the simulated experiments are designed and used for seven fiber positioning units neighbors.

This paper is organized as follows. In Sec. \ref{section2}, we analyze the collision probability in the overlapped area of the fiber units. Additionally, we explore the possible reasons for losing objects during the survey. The designation of hardware collision handling system is detailed in Sec. \ref{section3}. The results and discussions are summarized in Sec. \ref{sention4}.

\section{mechanical collisions between fiber positioning units }
\label{section2}
\subsection{The Fiber Allocation Rate }
 In a multi--fiber and dense distribution system, the allocation rate of the fibers should be taken into account.
Assuming that the distribution of objects in the focal surface of LAMOST is uniform, the probability (${p}$) of a object in a certain domain equals to the area ratio of the domain and the entire focal surface \citep{peng2003primary}. The probability of the objects distribution follows the binomial distribution which is shown in Eq. \ref{new}.
\begin{equation}
p(K,N)={C_{N}^{K}p^{K}(1-p)^{(N-K)}},
\label{new}
\end{equation}
where $N$ is the number of objects in the FoV, and $K$ is the number of objects in a domain.

\subsubsection{The Collision Probability of Two Objects in a Overlapped Area}
\begin{figure}
\begin{minipage}[h]{0.5\linewidth}
\centering
\includegraphics[scale=0.4,bb=0 0 600 350]{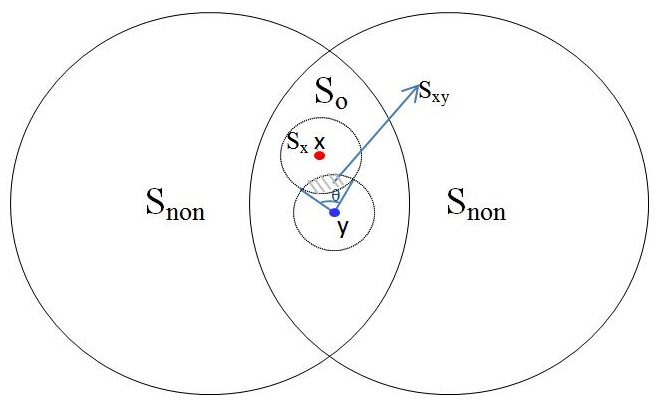}
\caption{Divided domains in a pair of fiber units. In this figure, $S_{o}$ represents the overlapped region, $S_{non}$ is the non--overlapped area, $x$ and $y$ are two targets, and $S_{x}$ is a circle area defined by the center $x$ and the radius 4.5 $mm$.}
\label{fig:side:x}
\end{minipage}%
\begin{minipage}[h]{0.5\linewidth}
\centering
\includegraphics[width=70mm,bb=0 0 600 350]{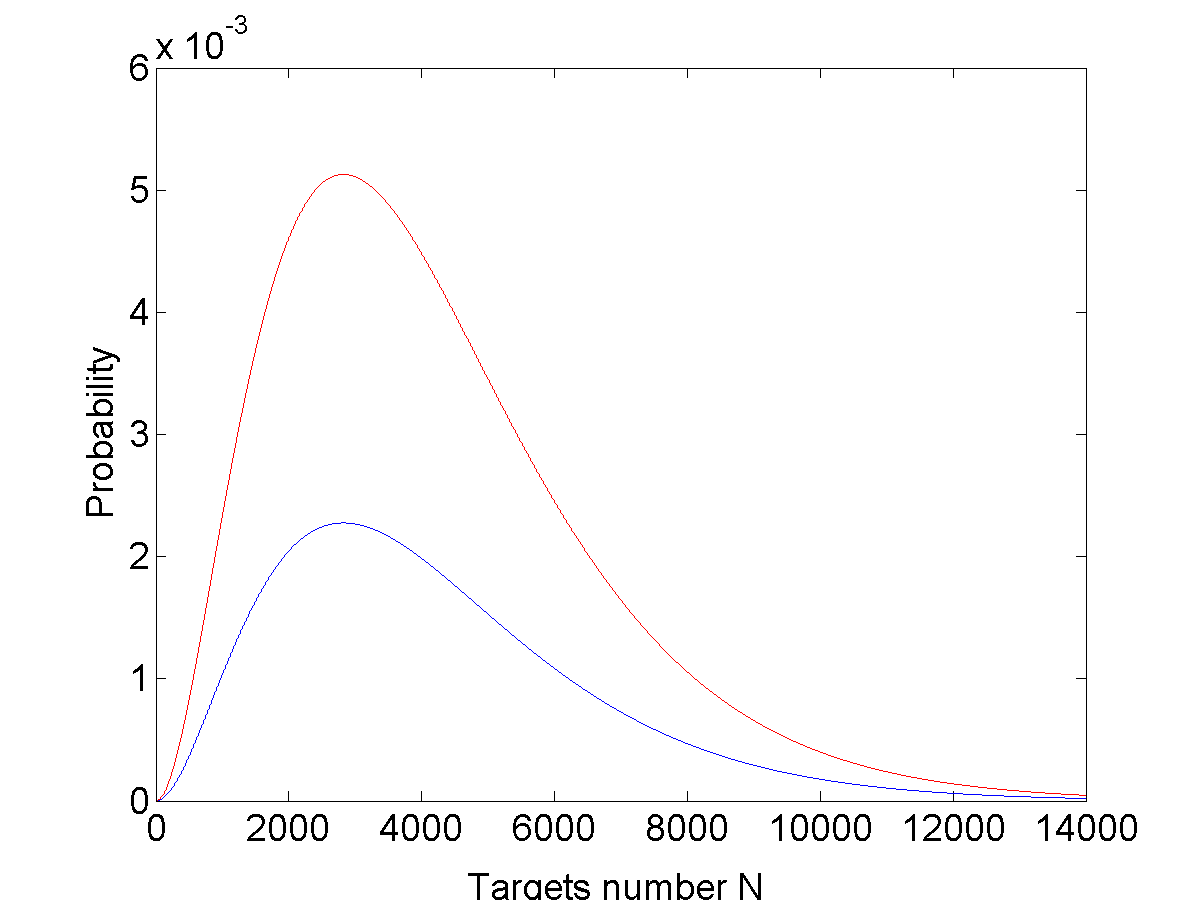}
\caption{The collision probability $p$ changes with $N$ (the number of objects). The red curve is the variation tendency of $p$ when the radius of $S_{x}$ is 6 $mm$ and the blue curve of which is 4.5 $mm$.}
\label{fig:side:y}
\end{minipage}
\end{figure}

Assuming that any pair of adjacent units is independent from the others, and each fiber unit has six neighbors, we estimate that about 11630 pairs of units are in the LAMOST fiber plate. A mechanical collision may happen in the overlapped region ($S_{o}$) as shown in Fig. \ref{fig:side:x}, since the fiber could not aim at the targets inside the structure of the fiber positioning-unit.
In the Fig. \ref{fig:side:x}, let  $x$ be a target in $S_{o}$, $S_{x}$ is the circle domain which center is $x$ and radius is 4.5 $mm$, the collision will occur when target $y$ falls in $S_{x}$. Thus, the collision probability of two objects in a overlapped area obeys the binomial distribution described as Eq. \ref{probability4}.

\begin{equation}
\label{probability4}
p={C_{N}^{2}p_{_{S_{o}}}^{2}(1-p_{_{S_{o}}})^{N-2}(1-p_{_{S_{non}}})^{N}p_{_{S_{x}}}},
\end{equation}
where $C_{N}^{2}p_{_{S_{o}}}^{2}(1-p_{_{S_{o}}})^{N-2}$ is the probability of only two objects in the overlapped region $S_{o}$, $C_{N}^{0}(1-p_{_{S_{non}}})^{N}$ is the probability of no objects in $S_{non}$, and $p_{_{S_{x}}}$ is the probability of the two objects allocated in $S_{x}$ simultaneously. We calculate this probability for different amount of targets listed in Table \ref{number}

The probability curve for the collision of the two objects is the $p$ vs. $N$ (various objects number) as plotted in Fig. \ref{fig:side:y}. The red curve represents $p$ changes along with $N$ when the radius of $S_{x}$ is 6 $mm$ and the blue curve of which is 4.5 $mm$. 4.5 $mm$ is just the theoretical radius of $S_{x}$, yet the radius is slightly larger than the theoretical one in practice. So we also considered 6 $mm$ as the radius of $S_{x}$ in the probability calculation, which would be a more reasonable value. In the curve shown in Fig. \ref{fig:side:y}, we see that the collision probability $p$ reaches a peak at the number about 3,500, which is close to the desirable number of targets (except for a few of skylight fiber units) in each observation. It is necessary to handle these collisions.

\subsubsection{The Collision Probability of More Objects in a Overlapped Area}
The collision probability caused by more than two objects allocated in the collision areas is relatively small. In the case of only three objects in $S_{o}$, the collision probability is defined as Eq. \ref{probability7} and its value is shown in Table \ref{number}.
\begin{equation}
p={C_{N}^{3}p_{_{S_{o}}}^{3}(1-p_{_{S_{o}}})^{N-3}(1-p_{_{S_{non}}})^{N}p_{_{S_{xy}}}},
\label{probability7}
\end{equation}
where $S_{xy}$ is the domain that defined by two objects $x$ and $y$, which is described as Eq. \ref{probability5} and the shadow part of Fig. \ref{fig:side:x}.
\begin{equation}
S_{xy}=\frac{{\int_{0}^{4.5}\frac{2[\frac{\theta\pi{r}^{2}}{2\pi}- {(r/2)}\sin{({\theta}/2)}l]2\pi{l}}{S_{o}}}dl}{\int_{0}^{4.5}2\pi{l}dl},
\label{probability5}
\end{equation}
where $l$$\in$ [0, 4.5] is the distance change of objects $x$ and $y$, $r$ is the radius 4.5 $mm$. Then the average area of $S_{xy}$ is 19.6539 $mm^{2}$ and the ratio of the possible collision zone in the fiber focal plane is $p_{_{S_{xy}}}$.

Generally, the total collision probability of any pair of fiber positioning units is defined as Eq. \ref{probability6}.
\begin{equation}
p=\sum_{j=2}^{N}C_{N}^{j}{p_{_{S_{o}}}^{j}(1-p_{_{S_{o}}})^{N-j}{(1-p_{_{S_{non}}})}^{N}{p_{_{S_{xy\dots{j}}}}}},
\label{probability6}
\end{equation}

In fact, $\sim$ 11630 pairs of fiber-positioning units are located on the focal plane of LAMOST. We select several groups of input targets number and calculate the relevant collision probability for the radius of 4.5 $mm$ and 6 $mm$ respectively. Table \ref{number} lists the result of $p_{_{S_{x}}}$, $p_{_{S_{xy}}}$, $n_{_{S_{x}}}$, $n_{_{S_{xy}}}$ and $\Sigma(n_{_{S_{x}}},n_{_{S_{xy}}})$. It suggests that the lost objects number tends to be the maximum when the input targets number of the FoV is about 3500. The amount of the lost targets reaches a hundred, therefore it is necessary to retrieve the lost targets considering the designation of LAMOST survey.

\begin {table}
 \scriptsize
 \caption{The Number of Lost Objects.}
 \label{number}
 \centering
 \begin{tabular}{cccccccccccccccccccccccccccccc}
\hline
\multirow{2}{*}{N} &
\multicolumn{2}{c}{2500} &
\multicolumn{2}{c}{3500} &
\multicolumn{2}{c}{6000} &
\multicolumn{2}{c}{8000} &
\multicolumn{2}{c}{10000} &
\multicolumn{2}{c}{14000} \\
\cline{2-13}
  & 4.5 mm & 6 mm & 4.5 mm & 6 mm & 4.5 mm & 6 mm & 4.5 mm & 6 mm & 4.5 mm & 6 mm & 4.5 mm & 6 mm \\
\hline
\multirow{2}{*}{$p_{_{S_{x}}}/n_{_{S_{x}}}$} &
 0.0022&0.005&0.0022&0.0049&0.0011&0.0024&0.0005&0.0011&0.0002&0.0004&0.0&0.0\\
\cline{2-13}
  &27 & 61 & 26 & 58 & 13 & 29 & 6 & 13 &2 &5 & 0 & 0\\
\hline
\multirow{2}{*}{$p_{_{S_{xy}}}/n_{_{S_{xy}}}$} &
0&0.0001 & 0.0001 &0.0002& 0.0001 &0.0002&0.0& 0.0 &0.0&0.0&0.0&0.0\\
\cline{2-13}
&0 & 6 & 8& 14 & 12 & 24 & 0 & 0 &0 &0 & 0 & 0\\
\hline
$\Sigma(n_{_{S_{x}}},n_{_{S_{xy}}})$&27&67&34&62&25&53&6&13&2&5&0&0\\
\hline
\end{tabular}
 \begin{flushleft}
 {\sc Notes:}\\
 1. $N$ means the input catalogue. 4.5 $mm$ is the theoretical minimal radius and 6 $mm$ is the more reasonable value in calculation of the collision probability.\\
 2. $p_{_{S_{x}}}/n_{_{S_{x}}}$ is the collision probability of the only two objects in the overlapped domain and the losing targets number respectively. $p_{_{S_{xy}}}/n_{_{S_{xy}}}$ is the ones of the only three objects in the overlapped domain and the losing targets number. \\
 3. The  probability of more than three targets in the overlapped domain in LAMOST is close to zero, so we don't list them in this table.\\

 4. $\Sigma(n_{2},n_{3})$ is the total number of losing targets of  $n_{2}$ and $n_{3}$.
 \end{flushleft}
\end {table}

\begin {table}
 \caption{Examples of Survey Strategies from LAMOST General Survey, where $M$ Represents the Number of Allocating Catalogue. }
 \label{survey strategies}
 \centering
 \begin{tabular}{cccccccccc}
 \hline

Plate&Time&	M&Observed Objects&Miss Objects&Failed Rate\\
\hline
GAC100N17B1&2014-12-26 23:17:30&3765& 3599&	166&4.4\%\\
HD053443N053939V01&	2015-01-28 19:30:00&3618&3522&96&2.6\% \\
GAC094N39M1&2015-02-23 20:20:06&3736&3638&98&2.5\%\\
HD103827N055449B02&2015-03-25 21:17:17&	3677&3536&141&3.8\% \\
HD145035N124702V01&2015-05-26 21:32:15&3492&3366&126&3.6\%\\

 \hline
 \end{tabular}
 \begin{flushleft}
 {\sc Notes:}\\
 This sample was selected from survey covering from 2014/12 to 2015/05. \\
 \end{flushleft}
\end {table}

\subsection{Discussion}
The collision probability reflects that the collision loss rate may reach about 3\%. While the average failed rate during the actual observations is as high as about 5\% as shown in Table \ref{survey strategies}, which lists five examples randomly selected from LAMOST observation results covering from 2014/12 to 2015/05. We infer that this loss rate difference may be induced by some other reasons. For example, some targets positioned by the fibers which are labeled as `abnormal' will be missed in the observation. The label `abnormal' means that these fibers have failed in locating for several positioning processes. The step motors for the fiber positioning units may miss some steps during LAMOST survey. Additionally, we do not exclude some operation mistakes.

Nevertheless, the lost objects number caused by the fiber units collisions reaches the level of 50 in every plate during the observations. It can bring down the efficiency of LAMOST to some extent. However, as many objects as possible should be observed since one of LAMOST scientific goals is to study the structure and the evolution of the Galaxy. It is valuable to retrieve these lost targets in the input catalogues. Handling the collisions between the fiber positioning units is very necessary. We design a hardware system to handle these situations during the observation stage, which will be described in the next section.


\section{handling collision system }
\label{section3}
\subsection{The Structure of the Hardware System}
\begin{figure}
\begin{minipage}[h]{0.5\linewidth}
\centering
\includegraphics[width=90mm,bb=0 0 500 350]{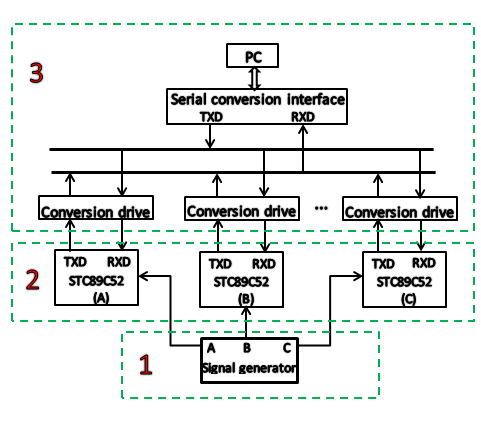}
\caption{The collision frame of hardware processing system. The label `1', `2' and `3' are a signal generation part, collision handling part and controlling system, respectively.}
\label{fig:side:b}
\end{minipage}
\begin{minipage}[h]{0.5\linewidth}
\centering
\includegraphics[width=80mm,bb=0 0 500 350]{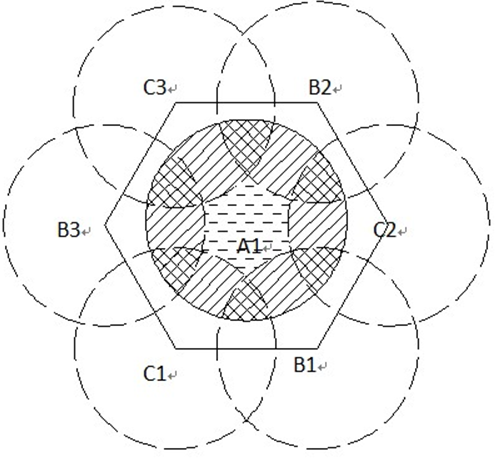}
\caption{Three types of collision signal schematic drawing.}
\label{fig:side:d}
\end{minipage}
\end{figure}

This hardware structure of the handling system is designed based on two programs which were built by the method of the voltage change inspection \citep{yan2007collision} and pulse contrast \citep{yan2008collision}. And it has been simulated successfully for one fiber positioning unit by the proteus, which is introduced in our series paper \citep{liuxiaojie2014simulations}. The hardware circuit structure of the handling collision includes three parts as shown in Fig. \ref{fig:side:b}.

The first is the signal generation part which is designed for detecting the collisions between the fiber positioning units labeled as `1'. In this part we choose the MSC51 series SCM as the signal controller. The second is the collisions handling part as labeled `2', which is the fiber units controller that gets the collision result based on the detection signal. The last part labeled as `3' is the microcomputer controlling system. In this system, a method based on VB and RS-232 serial interface\citep{zhao2006design} is utilized during the communications.

\subsection{The Designation of the Simulation Experiment}
I. \textit{Detection of Collisions Signal}\\
This part aims to detect the collisions and send the result to part `2'. The detection of collisions signal is based on the pulse superposition. To detect this superposition come from different fiber units, we should assign different pulses to the 4000 fiber units, theoretically. Because of the relative position of the adjacent fiber units, the overlapped region includes three units at most as seen in Fig.\ref{fig:side:d}. Therefore, we assign three types of pulses (A, B, C), which have the same frequency and different phases, as the detection signals for the 4000 fiber positioning units. This designation can assure any adjacent fiber signals are different.

The frequency of fiber units stepping motors is 510 $Hz$. To detect the collisions in real time, we apply a 200$\mu$s delay procedure by the MCU to generate the collision signals.When the collisions occur, the collision signals have multiple pulses in the same period.We choose the single chip STC89C52 as the core unit of pulse signal generator through the combination of the hardware circuit and software program phase. Therefore the frequency and amplitude of the waveform can change discretionarily in a certain range.

In the designation, firstly we set 1ms timer interrupt using the timer1 of STC89S52 chip as the time period of counting the collision pulses. Secondly during this period we count the collision pulses using counter0 of the chip. Thirdly when the timer1 interrupt occurs, if the value of the counter0 is 5, it suggests there is no collision. As a contrast, if the value is about between 10 and 15, we believe the fiber unit collisions with one or two adjacent units.

\noindent II. \textit{Signal processing}\\
This part describes how to process detection signals. It includes the host controller and the slave controllers. The host controller is to receive the information from the MCU and convey commands to the sub-controller (MCU). The main steps are detailed as follows.

\begin{figure}
\begin{minipage}[h]{0.5\linewidth}
\centering
\includegraphics[width=100mm,bb=-50 0 450 470]{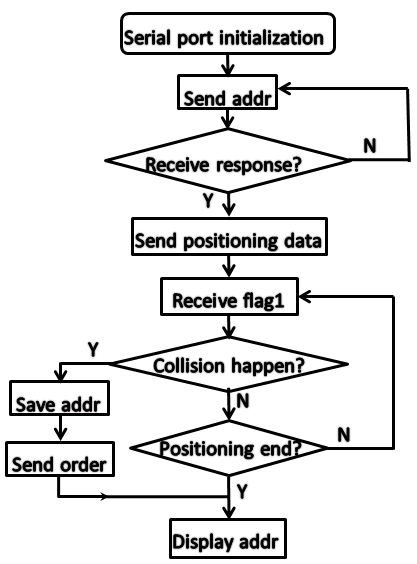}
\caption{The main controller of the collision handling system for LAMOST fiber positioning units process. The major function is sending the order to sub-controller and getting the situation of units.}
\label{fig:side:e}
\end{minipage}%
\begin{minipage}[h]{0.5\linewidth}
\centering
\includegraphics[width=100mm,bb=-50 0 450 470]{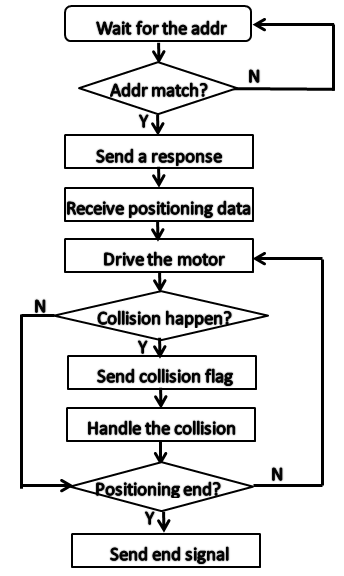}
\caption{The sub--controller of the collision handling system for LAMOST fiber positioning units process. The program includes receiving and feeding back the information of controlling units from the main controller.}
\label{fig:side:f}
\end{minipage}
\end{figure}

When the subsystem based on MCU sends the collision results to microcomputer through the serial bus, the main program on microcomputer receives the results and determines actual position of the fiber units timely. As a result, microcomputer can timely detect the fiber position units survey situation just through the data received from sub-controller. The Fig. \ref{fig:side:e} and Fig. \ref{fig:side:f} explain the procedure of the serial data communication including serial data reception and transmission.

The sub-controllers implement two functions. One is to receive the collision results and transfer them to microcomputer controlling system, the other is to execute the instructions from microcomputer. STC89C52 MCU produced by company ATMEL \citep{liu2012mcu} is used to control the positive inversion of stepping motor.  The collision results will be sent to microcomputer by MSComm. The MCU will control the rotation way of units stepping  motors when the instructions coming\citep{liuxiaojie2014simulations}.

\noindent III. \textit{Serial communication}\\
The serial communication between microcomputer and single chips plays a key role in achieving the above functions. By using MSComm control, microcomputer series-communicating with many MCUs controls the fiber unit motors steps and returns collision event information automatically. The whole hardware circuit consists of the slave MCUs which are considered as the second level systems and microcomputer. The MCU serial communication at a TTL level remains between 0 V and 5 V, while the microcomputer serial communication stands for the RS-232 standard. The most feasible solution to connect these two serial ports is just plugging a MAX-232 between them as the serial conversing interface\citep{ege2013new}. Thus, the steps of unit motor rotating are calculated by microcomputer according to the position of the objects and transmitted through the serial bus\citep{tian2007design}, which connects the microcomputer with the sub-controller. During the implement, one or two standard serial ports (COM1 or COM2), aims at achieving serial data communication through VB(Visual Basic) programming.

\subsection{The Result Analysis}
These three modules run orderly according to the modules function. When the collisions  occur between the adjacent fiber positioning units during the observation, the hardware system is effective to handle this situation. The signal generator can be counted by the sub-controller without AD converter, and it also has the advantages of small volume, low price, stable performance and complete function. The serial communication method has characteristics of simple connection, high reliability, low cost, etc.
This simulated experiment is for seven close neighbor fiber units. When this system is put into use for 4000 fiber units, the sub - controller MCU should be changed with FPGA which has high integration character. So one chip of FPGA can control more fiber units.
\section{summary}
\label{sention4}
In this work, we first analyze the collision probability between fiber positioning units. To get this probability, we  assume that the allocation of the targets in the focal plan is uniform, so it obeys the binomial distribution. Table \ref{number} shows two boundaries under the radius of $S_{x}$ $r$ = 4.5 $mm$ and $r$ = 6 $mm$. In fact, the most possible collision may happen at the radius ~5 $mm$ (between 4.5 $mm$ and 6 $mm$) of $S_{x}$. If we choose the radius as 5 $mm$, so  the possible collision number of the fiber units  will be about 67 when the number of the targets allocated in the focal surface is about 3500. It is worthy of finding a way to handle these fiber units collisions.
We developed a new real-time hardware handling collision system for LAMOST fiber positioning units. It is a master - slave control system. In the designation we use microcomputer as our host controller which sends the position data and the collisions handling commands to the sub - controllers. The STC89C52 single - chip computer is used as slave controller which receives the host controller located commands and accepts the data from feedback.

Through this system, the broken fiber units could be marked accurately and changed efficiently. Additionally, it can improve the efficiency of LAMOST fiber units and the completeness of the survey. Because LAMOST has been operated for several years, some fiber units should be repaired or replaced at regular intervals. To a certain extent, the hardware system may also lengthen the service time of the fiber units. It is a complex system for handling 4000 fiber units, further research is required in the future.

\begin{acknowledgements}
First and foremost,  the authors would like to show our deepest gratitude to our colleague, who have provided us with valuable guidance in every stage of the writing of this thesis. Without their enlightening instruction, impressive kindness and patience, this paper could not have been completed.

Guoshoujing Telescope (the Large Sky Area Multi-Object Fiber Spectroscopic Telescope LAMOST) is a National Major Scientific Project built by the Chinese Academy of Sciences. Funding for the project has been provided by the National Development and Reform Commission. LAMOST is operated and managed by the National Astronomical Observatories, Chinese Academy of Sciences.
\end{acknowledgements}

\bibliographystyle{raa}
\bibliography{bibtex}

\begin{thebibliography}{12}
\providecommand{\natexlab}[1]{#1}
\providecommand{\selectlanguage}[1]{\relax}

\bibitem[{{Chen} et~al.(2014){Chen}, {Bai}, {Luo}, \&
  {Zhao}}]{2014SPIE.9149E..1NC}
{Chen}, J.-J., {Bai}, Z.-R., {Luo}, A.-L., \& {Zhao}, Y.-H. 2014, in Society of
  Photo-Optical Instrumentation Engineers (SPIE) Conference Series,
  \emph{Society of Photo-Optical Instrumentation Engineers (SPIE) Conference
  Series}, vol. 9149, 1

\bibitem[{Cui et~al.(2012)Cui, Zhao, Chu et~al.}]{cui2012large}
Cui, X.-Q., Zhao, Y.-H., Chu, Y.-Q., et~al. 2012, Research in Astronomy and
  Astrophysics, 12, 1197

\bibitem[{Ege et~al.(2013)Ege, {\c{S}}ensoy, Kalender, Nazl{\i}bilek, \&
  {\c{C}}{\i}tak}]{ege2013new}
Ege, Y., {\c{S}}ensoy, M.~G., Kalender, O., Nazl{\i}bilek, S., \&
  {\c{C}}{\i}tak, H. 2013, Measurement, 46, 2672

\bibitem[{Li et~al.(2000)Li, Hu, \& Yu}]{Liweimin+2000}
Li, W., Hu, H., \& Yu, Q. 2000, Instrument Technique and Sensor, 4, 33

\bibitem[{Liu et~al.(2012)Liu, Sun, Zhao, Yao, \& Zhang}]{liu2012mcu}
Liu, J., Sun, G., Zhao, D., Yao, X., \& Zhang, Y. 2012, Procedia Engineering,
  29, 2109

\bibitem[{Liu \& Wang(2014)}]{liuxiaojie2014simulations}
Liu, X., \& Wang, G. 2014, J.USTC, 11, 423

\bibitem[{Peng et~al.(2003)Peng, Zhai, \& Xing}]{peng2003primary}
Peng, X.-b., Zhai, C., \& Xing, X.-z. 2003, JOURNAL-UNIVERSITY OF SCIENCE AND
  TECHNOLOGY OF CHINA, 33, 78

\bibitem[{Tian et~al.(2007)Tian, Zhao, \& Zeng}]{tian2007design}
Tian, Y.-j., Zhao, G.-q., \& Zeng, J.-p. 2007, Journal of Hunan Institute of
  Engineering (Natural Science Edition), 2, 006

\bibitem[{Xing et~al.(1998)Xing, Zhai, Du et~al.}]{xing1998parallel}
Xing, X., Zhai, C., Du, H., et~al. 1998, in Astronomical Telescopes \&
  Instrumentation, 839--849 (International Society for Optics and Photonics)

\bibitem[{Yan et~al.(2007)Yan, Zhai, \& Shu}]{yan2007collision}
Yan, F., Zhai, C., \& Shu, Z. 2007, Machinery \& Electronics, 08, 3

\bibitem[{Yan et~al.(2008)Yan, Zhai, Shu et~al.}]{yan2008collision}
Yan, F., Zhai, C., Shu, Z., et~al. 2008, Chinese Journal of Scientific
  Instrument, 8, 014

\bibitem[{Zhao et~al.(2006)Zhao, Yuan, \& Yang}]{zhao2006design}
Zhao, J., Yuan, Z.-f., \& Yang, C.-s. 2006, China Meas Urement, 6, 030

\end{thebibliography}
\end{document}